# Unit Commitment Considering the Impact of Deep Cycling

HyungSeon Oh

Department of Electrical and Computer Engineering, United State Naval Academy, Annapolis MD 20401

Email: hoh@usna.edu

*Abstract* – **Wind energy has been integrated into the power system with the hope that it improves the energy efficiency and decreases greenhouse gas emission. However, several studies over the world imply that the result was in the opposite way that was hoped mainly because of the negative correlation between wind availability and load. Under the situation, coal power plants are forced to cycle while they are not designed to do so. To prevent this unwanted result from occurring, a unit commitment decision should include the use of fuel and the emission rate during the ramp up/down process. This paper proposes a new unit commitment decision process to accommodate the economic and the environmental costs associated with the ramping process. The costs are, in general, not convex because there is positive cost if a generator output changes significantly regardless of directions. As a result, the problem might be nonconvex. A piece-wise linear cost curve is introduced to model the impact of ramping processes. With the curve, a convex linear programming is formulated, and the impact of a governmental policy is discussed.**

## I. NOMENCLATURE

| | |
|---|---|
| $B_{branch}$ | Branch impedance matrix |
| $B_{bus}$ | Bus impedance matrix |
| $GA$ | Set of generators with high ramp rates |
| $GB$ | Set of generators with low ramp rates |
| $H$ | PTDF matrix with cardinality of $L$-by-$N$ |
| $K_-$ | Diagonal matrix with minimum generation |
| $K_+$ | Diagonal matrix with maximum generation |
| $L$ | Number of lines in a transmission network |
| $L_{GA}$ | Location matrix of generators in $GA$ |
| $L_{GB}$ | Location matrix of generators in $GB$ |
| $LL$ | Location matrix of storage devices |
| $L_g$ | Matrix indicating the location of generators with the cardinality of $N_g$-by-$N$ |
| $N$ | Number of buses in a transmission network |
| $NC$ | Diagonal matrix with no-load cost |
| $N_g$ | Number of generators |
| $R$ | Maximum ramp rate vector |
| $RU$ | Combined cost vector of the impact on the environmental and the life time of the generator during its ramp up process |
| $RD$ | Combined cost vector of the impact on the environmental and the life time of the generator during its ramp down process |
| $SP$ | Diagonal matrix with maximum spinning reserve |
| $SU$ | Diagonal matrix with start-up cost |
| $d_t$ | Expected load at time $t$ |
| $diag(x)$ | Diagonal matrix with the diagonal elements of $x$ |
| $e_j$ | $j^{th}$ unit vector |
| $g$ | Power generation vector with the cardinality of $N_g$-by-$1$ |
| $\Omega$ | Hourly recovery factor of the construction cost for a storage device |
| $\alpha$ | Ramp up variable |
| $\beta$ | Ramp down variable |
| $\delta_{t,j}$ | Energy stored in the storage device $j$ at time $t$ |
| $\varepsilon_C$ | Charging efficiency of a storage devoice |
| $\varepsilon_D$ | Discharging efficiency of a storage device |
| $\kappa$ | Ratio of economic minimum to the maximum generation of the unit of interest |
| $\rho$ | Spinning reserve margin |
| $\gamma$ | Charging variable of a storage device |
| $\theta$ | $N$-by-1 voltage angle vector |
| $\tau$ | time for ramping up/down |
| $\tau_h^i$ | Minimum down-time of generator $i$ |
| $\tau_s^i$ | Minimum up-time of generator $i$ |
| $v$ | Discharging variables of a storage device |
| $\xi$ | Energy rating of storage devices |
| $\psi$ | Power rating of storage devices |

## II. INTRODUCTION

Recent ten years or so time, the concern of global warming has been raised. In the effort to reduce the greenhouse gas, many wind farms have been constructed in many countries. Beside the low emission rates, they would bring the electricity price down due to their low operation costs. However, recent studies have shown that the greenhouse gas emission increases and that the system cost increases [1], [2]. In some circumstances, the negative price bidding on the purpose of getting dispatched by renewable sources for tax credits may increase power system emissions [3]. A most important reason is that the availability of wind does not positively correlate with demand [4], [5]. For example, early in the morning, load is low while wind resource is abundant, and the situation becomes reversed around noon. Different from wind resources, solar energy tends to be positively correlated to the loads, and therefore the integration of wind energy may not result in the dispatch changes of coal power if the solar or other energy resources mitigate the impact of the wind energy. One way to mitigate the impact of the negative correlation would be to integrate storage devices [6], showing the reduction of emissions as power systems increase the generation portfolios of low- and no-carbon generation sources [7].

An objective of the steady-state operation of power systems is to fulfill the loads continuously regardless of the temporal variation of loads and generation. A unit commitment problem (UCP) is an optimization that includes multi-period optimal power flow with start-up and shut-down options of multiple generators, which are binary variables. Due to the large scale of power systems, a conventional UCP includes a linear optimal power flow. Therefore, the UCP is classified as a mixed integer linear programming. In UCP, an hourly schedule of which generators are on is determined to meet load in a lowest cost. During a peak period, a high level of generation is required to meet the load, and therefore, many generators are committed. Some generators must be ready ahead of dispatch time due to their long minimum-down-time and/or long maximum-ramp-rates [8]. It is possible to ramp up at its maximum ramp rate starting from an uneconomic operation point for coal power plants, but such an operation leaves an negative impacts on the generators' life time – dynamic impact [5]. Since ramping under a certain generation level of coal power plants has a significant emission and economic impact, it is difficult to incorporate the impacts in a conventional UCP. Any change in dispatches between two periods of a generator is allowed with no penalty as long as feasible. For the operation with only conventional generators, the temporal changes in loads is slowly varying and therefore the dynamic impacts are not imposed frequently. The high penetration level of wind energy, however, changes the situation due to the characteristics of wind energy resources; 1) abrupt change in the resource in a relatively short time and 2) the negative correlation between the energy resource and the demand. Consider a system with no storage devices (or highly limited capacity) due to their high costs. It is possible for the wind energy resources to change from abundant to low within an hour that is a typical time difference between two time slices in UCP while loads in the duration remain low. At the first time slice (low loads but high wind), the generators with high operation costs (peakers) are forced to shut down. Since winds are generated at very low generation costs if not nonzero, wind generations are fully dispatched, and the difference between loads and wind powers should be satisfied by the base units. At the second time slice, the wind generation drops significantly while loads stay constant, the changes in wind resource must be compensated by the fuel-burning generators. The generators that are turned off at the first time slice can produce their output no more than their maximum ramp rates only if they have zero minimum-down-time, which may not be sufficient to compensate the change in wind energy resources. The base units should ramp up to meet the loads. In general, the base units have a low (or practically zero) ramp rate except for hydropower plants that have highly constrained in terms of their operations to meet fishing or leisure demands. Among base units, coal power plants are least constrained in terms of the change in their outputs, and therefore forced to change the output in this situation. Therefore, with increasing the penetration level of wind energy, coal power plants are also forced to change their generation outputs significantly (cycling) [1]. Different from gas power plants (peakers), coal plants are not designed for cycling [5]. As a result of the coal cycling, the emission rates increase, and the life of the coal plants system reduces. There have been attempts to control the environmentally toxic gases in the power system operation [9-16]. However, the ramping impact on the emission from coal plants is not properly addressed.

To avoid coal cycling; (1) one may reduce the ramp rates of coal plants to avoid a significant change in their generation outputs, and/or (2) the impacts of ramping on the emission and the cost due to the reduced plant life are added in the system cost that is the objective function of the UCP. The former may overly constrain the system and may result in infeasible solution while the latter finds one. In order to include the additional costs in mixed integer linear programming, they should be convex. However, both the emission rates and the negative impact on the plant life increase as the change in output regardless of the direction of the change. Therefore, the variable to express the change in generation is not convex. In this paper, two variables are introduced instead, and convex mixed integer linear programming is formulated and tested in a model system.

This paper is organized as follow: Section III models the ramping impact on the greenhouse gas emission; Section IV models the damage of coal cycling; Section V presents an optimization problem to address the coal cycling; Section VI presents the numerical results; and Section VII lists conclusions and future research directions.

## III. DYNAMIC EMISSION MODELING

In this study, dynamic emission terms the additional $CO_2$ emission while a power plant undergoes the change in its output. A yearly emission data of three coal power plants are provided by *Xcel Energy*. The data set is comprised of the measured $CO_2$ emission data in each hour. Suppose at time $t$, the generation is $g$, and the output before and after time $t$ are $g_0$ and $g_1$, and the generator can ramp up/down its output in $\omega g_{max}$ in an hour. Typically coal power plants have the capability to change its output $1.5 - 5\%$ of $g_{max}$ in a minute [4]. Therefore, $\omega$ is in the range of $0.9 - 3$. With the transition time $\tau$, the ramp rates $\omega_0$ from $g_{t-1}$ to $g_t$ and $\omega_1$ from $g_t$ to $g_{t+1}$ are defined:

$$\omega_0 = \frac{|g_t - g_{t-1}|}{\tau g_{max}} \text{ and } \omega_1 = \frac{|g_{t+1} - g_t|}{\tau g_{max}} \text{ where } \tau \ll 1 \qquad (1)$$

Note that the transition time should be less than an hour because the generations are dispatched every hour. This transition times are split into two pieces and assigned to previous and precedent time periods, i.e., the last $\tau/2$ hours from the *(t-1)* period and the first $\tau/2$ hours from the *t* period are spent to change the output from $g_0$ and $g$, and similar assignment is performed between $t$ and *(t+1)* periods. In such a case, we have three periods of the generation outputs, transition from $g_{t-1}$ to $g_t$, constant generation at $g_t$, and transition from $g_t$ and $g_{t+1}$. Correspondingly, there can be three periods of emission of $CO_2$, static emission.

A coal power plant is typically a base unit, and therefore its generation is fixed for its most operation. The generation data are collected for a further analysis only when the plant does not ramp up/down for at least three consecutive hours. We propose two incremental emission models when a coal power plant generates a fixed output for multiple periods; a polynomial and a linear model.

$$f(g) = f_0 + f_1 g_t^{N_1} \tag{2}$$

where $f$ is an incremental emission function at a given generation $g$. Note that (2) is generalized by introducing a piece-wise incremental model, i.e., for a range of generation output, the parameters ($f_0, f_1$, and $N_1$) are assigned. An hourly change in generation for a coal power plant is limited, and therefore, we assume the change occurs within a same range so that (2) holds with a single set of the parameters. Because the emission data are collected in an hour, the measured emission data is the integration of the instantaneous emission function over an hour, i.e.,

$$F(g) = \int_0^1 f(g) dt = \int_0^1 f_0 + f_1 g_t^{N_1} dt = f_0 + f_1 g_t^{N_1} \tag{3}$$

The fitting result with the polynomial model yields the value of the slop $N_1 = 0.9460$. It is possible to develop a model to allow the change in generation to cover multiple ranges (multiple sets of the parameters for (2) and (3)) and to perform a segmented regression [17]. Figure 1 shows the emission data when the coal power plant generates a fixed output. Over the entire generation range, a single line fits the data well. We found no evident improvement of fitting results from segmented regression over a single region, and therefore, in this study, one set of parameters will be used. The parameters of the polynomial model $f_0, f_1$, and $N_1$ are determined from the linear regression to (3) 11.53 ± 2.37, 0.86 ± 0.09, and 1.02 ± 0.02, respectively.

Figures 2 and 3 present the $CO_2$ emission when the same coal power plant undergoes a change in its generation. We first show the $CO_2$ emission in terms of the generation output by ignoring the dynamic emission in Figure 2. While the line fits the data well in the range of above 50% of the maximum generation, the data are above the line below 50%. This is a clear evidence of the impact on the dynamic emission. Figure 3 demonstrates the impact of the dynamic emission that the estimated emission from both static and dynamic emission. The static emission is computed based on (3), and the impact of the dynamic emission is discussed below.

As is in (1), when a coal power plant undergoes the change its output, the change in output occurs in a fixed time $\tau$ regardless the change in the model. In the case, the power generation in an hour may have three different generation stages, i.e., $g_I$ (transition from $g_{t-1}$ and $g_t$ in $[-\tau/2, \tau/2]$), $g_t$ (fixed generation in $[\tau/2, 1-\tau/2]$), and $g_{II}$, (transition from $g_t$ to $g_{t+1}$ in $[1-\tau/2, 1+\tau/2]$).

$$\begin{cases} g_I = \frac{g_{t-1} + g_t}{2} + \omega_I t = \frac{g_{t-1} + g_t}{2} + \frac{g_t - g_{t-1}}{\tau} t \\ g_{II} = \frac{g_t + g_{t+1}}{2} + \omega_{II} t = \frac{g_t + g_{t+1}}{2} + \frac{g_{t+1} - g_t}{\tau} t \end{cases} \tag{4}$$

In the consideration of the emission over an hour, only instantaneous emission within an hour is taken into account. For the instantaneous emission during the change in generation output, we propose a model as follows;

$$f(g,\omega) = f_0 + f_1 g_t^{N_1} + 2b(\Delta g)^{N_2} \tag{5}$$

Note that (5) is consistent with the linear model in (2) because in a static case $\Delta g$ becomes zero.

$$\begin{aligned} F(g,\Delta g) &= \int_0^{\tau/2} f(g)dt + \int_{\tau/2}^{1-\tau/2} f(g)dt + \int_{1-\tau/2}^{1} f(g)dt \\ &= \int_0^{\tau/2} \left[ f_0 + f_1 g_t^{N_1} + 2b(\Delta g)_I^{N_2} \right] dt + \int_{\tau/2}^{1-\tau/2} f_0 + f_1 g_t^{N_1} dt + \int_{1-\tau/2}^{1} f_0 + f_1 g_t^{N_1} + 2b(\Delta g)_{II}^{N_2} dt \\ &= F_0(g_{t-1},g_t,g_{t+1},\tau) + b\tau \left( |g_t - g_{t-1}|^{N_2} + |g_{t+1} - g_t|^{N_2} \right) \end{aligned} \tag{6}$$

where $F_0(g_{t-1},g_t,g_{t+1},\tau) = f_0 + f_1 g_t^{N_1} + \dfrac{f_1\tau}{N1+1} \left\{ -g_t^{N_1}(N_1+1) + \dfrac{\left(\dfrac{g_{t+1}+g_t}{2}\right)^{N_1+1} - g_t^{N_1+1}}{g_{t+1}-g_t} + \dfrac{g_t^{N_1+1} - \left(\dfrac{g_t+g_{t-1}}{2}\right)^{N_1+1}}{g_t - g_{t-1}} \right\}$.

Even a base unit is designed to change its output, and therefore, such dynamic effect on emission may not be evident on certain range in the operation. Because a typical base unit operates near its full capacity, the effect can be limited at a low generation. The emission data obtained when a coal power plant changes its output are plotted in Figure 2.

In Figure 3, the dotted line indicates the expected emission of $CO_2$ if the plant generates a fixed output – static emission. As is shown, the dynamic emission becomes evident when the output is less than economic minimum generation, $\kappa \times$ maximum generation, which leads to the underestimation of $CO_2$ emission using the static model with the best-fit result with (3) below 300 MW, $\kappa = 0.5$. To consider the dynamic emission that is the second term in (6), the $CO_2$ emission data at the generations below 300 MW are fitted with a nonlinear equation in (6). Figure 3 illustrates the best-fit results. The transition time is assumed 10 minutes. The blue line in Figure 2 indicates the estimated $CO_2$ emission with the consideration of dynamic impact on the emission of the coal cycling.

Because this dynamic emission is evident when a coal power plant operates at a low generation, a coal power plant is modeled as two plants – with and without dynamic emission. Two generators are completely independent except the fact that one with dynamic emission must be dispatched before the other one generates electricity. This condition is implemented with an additional constraint, i.e.,

$$g_{max}^{m,I} u_t^{m,II} \le g_t^{m,I} \quad \forall t \to \begin{cases} u_t^{m,II} = 0 \text{ if } g_t^{m,I} > g_{max}^{m,I} \\ u_t^{m,II} = 0 \text{ or } 1 \text{ if } g_t^{m,I} = g_{max}^{m,I} \end{cases} \tag{7}$$

where $I$ and $II$ represent the units with and without dynamic emission effect from a coal power plant m, respectively. (7) prevents the situation that $II$ can be dispatched only after $I$ runs its maximum capacity. Note that in the model total emission from $II$ is greater than that from $I$ even without the dynamic emission effect, and therefore, the objective function is still convex after taking the emission into consideration. This analysis is also performed for gas power plants, but the value for $\omega$ is very large, and $\tau$ is negligible. Therefore, in this study, the dynamic emission rate is only assumed for a coal power plant.

## IV. DAMAGE OF COAL CYCLING

Coal power plants have traditionally played a role of base unit, i.e., most time they operate in their full capacities. Therefore, they were not designed to change their output frequently. If they do, it is believed to have a negative impact on their life. There are several studies performed to evaluate the impact of coal cycling on the life shortening. To model the damage on the coal power plant, startup penalty is the most significant factor [4], and its impact can be implemented in terms of the startup cost. For example, for the coal plant in Section III, the cost is related to the first unit of the plant *I*.

Deep cycle refers the incidence that a coal power plant generates electricity less than its economic minimum. In the current unit commitment tool, deep cycle is not taken into consideration; however, deep cycle is an important factor to contribute the increase in cost. Various reasons are mentioned why the operation of a coal power plant generates below its economic minimum increases the cost [8]. In Section III, a coal power plant is modeled with two units, *I* and *II*. Deep cycling is equivalent to the ramping up/down of the unit *I*. As a result, we propose a new unit commitment tool that includes two different dynamics for a coal power plant.

IV.1. Ramping up and down

The system operation does not involve any penalty for the change in dispatch between *t-1* and *t* as long as the change is within the feasible range of the ramp rates. However, on top of the additional $CO_2$ emission, recent studies show that significantly large changes in dispatch performed below a certain level of generation may result in negative impact on the lifetime of coal power plants and on greenhouse gas emission [8]. In this study, the level is termed an economic operation level (EOL) and such impacts below EOL are taken into consideration. There are two directions of the changes in the generation output; ramp-up, $\alpha_t$, and ramp-down, $\beta_t$, i.e.,

$$\begin{cases} \alpha_t \equiv g_t - g_{t-1}, \beta_t \equiv 0 & \text{if } g_t > g_{t-1} \\ \alpha_t \equiv 0, \beta_t \equiv -g_t + g_{t-1} & \text{if } g_t < g_{t-1} \\ \alpha_t \equiv 0, \beta_t \equiv 0 & \text{if } g_t = g_{t-1} \end{cases} \quad (8)$$

(8) is equivalent to $\alpha_t \geq g_t - g_{t-1}, \beta_t \geq -g_t + g_{t-1}$ where $\alpha_t \geq 0, \beta_t \geq 0$ if $\alpha_t$ and $\beta_t$ are forced to be minimized since (8) is the result of any convex function of $\alpha_t$ and $\beta_t$ in the feasible region defined by the equivalent conditions. Figure 4 illustrates how $\alpha_t$ and $\beta_t$ are defined in terms of the change in generation output in two adjacent time between *t-1* and *t*. Suppose the output of a generator *i* at time *t* increases by 10MW, which yields $\alpha_t^i = 10$ and $\beta_t^i = 0$. It is clear that either $\alpha_t^i$ or $\beta_t^i$ are zero unless both are zeros.

The impact of ramping is two-fold; the greenhouse gas emission and the impact on the power plant life. A way to include the greenhouse emission in the objective function is to charge on the emission such as carbon tax. This environmental cost associated with the greenhouse gas emission enters into the operation cost *c* as well as the dynamic emission, i.e., the cost is associated with the generation as well as the change in generation. From Section III, the impact on dynamic emission is estimated using (6). In a similar way to break up the operation cost with a step function in an offers, we propose a step function to incorporate the dynamic emission, i.e., for a block of $\alpha$ or $\beta$ between $[\Delta_1, \Delta_2]$, the representative dynamic emission is assigned to make the area under the actual and the step function equal:

$$\int_{\Delta_1}^{\Delta_2} F(g, \Delta g) d\alpha = \frac{b\tau}{N2+1}\left(\Delta_2^{N_2+1} - \Delta_1^{N_2+1}\right) = (\Delta_2 - \Delta_1)\overline{F} \rightarrow \overline{F} = \frac{b\tau}{N2+1}\frac{\Delta_2^{N_2+1} - \Delta_1^{N_2+1}}{\Delta_2 - \Delta_1} \tag{9}$$

Even if (9) is derived for the case when the ramp rate increases ($\alpha > 0$ and $\beta = 0$), the representative value $\left(\overline{F}\right)$ can also be estimated in a similar way. While the estimated dynamic $CO_2$ emission varies continuous with the change in generations (blue line), it is desirable to establish a step function to construct an MILP for UCP. Figure 5 illustrates how the step function is constructed. The step function should represent the estimated dynamic emission of $CO_2$, and the area under the curve (cumulative emission during the time periods). The condition should hold either $\alpha$ (ramp up; $\alpha > 0, \beta = 0$), or $\beta$ (ramp down; $\alpha = 0, \beta > 0$). The impact on the power plant life is also affected by the cycling unless the plant was designed to do so. Ref [4] indicates that both impacts increase with increasing $\alpha_t$ and $\beta_t$. Because the damage from the coal cycling is to operate coal power plants below certain output (deep cycle), the cost is associated with shutdown of a second part of the coal unit.

Note that the values for RU and RD are strictly positive for non-zero $\alpha_t$ and $\beta_t$. Because the feasible regions for $\alpha_t$ and for $\beta_t$ from the constraints are above the linear constraints and they are not involved with other constraints, the minimization process forces $\alpha_t$ and $\beta_t$ to stay on the lines because the deviation of $\alpha_t$ and $\beta_t$ from the equality constraints in (8) results in the increase in the total cost. Therefore, the inequality constraints of $\alpha_t$ and $\beta_t$ are introduced in terms of the equality constraints shown in (8). As Figure 6 illustrates how the optimization process that forces $\alpha_t$ and $\beta_t$ move on to the line identical to equality constraints shown in (8), the optimization process forces the inequality constraint associated with $\alpha_t$ and $\beta_t$ identical to (8).

IV.2. Sequencing of two units to represent a single coal power plant

In the UCP, for each unit model of a coal power plant, generation, unit commitment, startup, and shutdown variables are introduced. The internal relationship should be established to represent the operation of a single generator. For example, there is a sequencing constraint since Unit II is on only if Unit I is fully dispatched. Suppose there is a coal power plant whose EOL and generation capacity are 50 and 100 MW, respectively. If the generation is 47 MW, Unit I operates at 47 MW while Unit II is off. This requires $\overline{g}_m^{EOL} u_t^{m,II} - g_t^{m,I} \leq 0$ where $\overline{g}_m^{EOL}, u_t^{m,II}$, and $g_t^{m,I}$ are the maximum EOL of the $m^{th}$ coal power plant, unit commitment variable of Unit II, and dispatch variable of Unit I, respectively. A commitment order constraint exists that Unit II is on under a condition that Unit I is on, $u_t^{m,II} \leq u_t^{m,I}$. Note that the sequencing constraint includes the commitment order constraint since $\overline{g}_m^{EOL} u_t^{m,II} - g_t^{m,I} \leq 0$ combined with $g_t^{m,I} \leq \overline{g}_m^{EOL} u_t^{m,I}$ trivially satisfies the commitment order constraint. In addition, the total generation from Unit I and II equals the dispatch from the corresponding coal power plant, i.e., $g_t^{m,I} + g_t^{m,II} = g_t^m$. The startup and the shutdown variables of both units are not associated with any costs and any constraints if the variables are introduced for the corresponding coal power plant. Therefore, the startup and the shutdown variables of the units are not introduced.

## V. PROBLEM FORMULATION

Ref [18] outlines a centralized unit commitment to determine 24 hurly dispatches in day-ahead markets.

V.1. Variables

The control variables $x$ of the UCP in this study are the voltage angle $\theta$; generation $g_t$; generations from two units of coal power plants $g_t^{coal,I}$ and $g_t^{coal\,II}$; unit commitment variable $u_t$; unit commitment variable of two units of coal power plants $u_t^{coal,I}$ and $u_t^{coal\,II}$; start-up variable $s_t$; shut-down variable $h_t$; ramping up variable $\alpha_t$ of Unit I; ramping down variable $\beta_t$ of Unit I; cost variable associated with generation $y_t$; cost variable associated with ramping up of Unit I $y_t^\alpha$; cost variable associated with ramping down of Unit I $y_t^\beta$; charge rate $\gamma_t$; discharge rate $\nu_t$; charge state $\delta_t$ for multiple scenarios $k$'s at each time $t$. Note that the Unit I and II are associated with the coal power plants only.

V.2. Inputs

The inputs for the UCP are the network parameters are introduced in terms of $B_{bus}$, $B_{br}$ matrices, and line capacities; the locational loads over the time span under consideration; the maximum ramp rates $R$; the generation limits $K_+$ and $K_-$; the EOL limits of Unit I for the $m^{th}$ coal power plant $\bar{g}_m^{EOL}$; the minimum uptime $\tau_s$ and the minimum downtime $\tau_h$ for the generators; the histories that indicate when the startup and the shutdown occurs last time; the cost curves associated with $g_t$, $\alpha_t$, and $\beta_t$; no-load costs for the generator $NC$; the characteristics of storage devices are rated power $\psi$, rated energy $\xi$, charging efficiency $\varepsilon_C$, discharging efficiency $\varepsilon_D$; and the duration of a time slice $\lambda_t$.

V.3. Objective function, $\Phi$

The objective function comprises 1) generation costs, 2) start-up cost, 3) deep cycling cost to recover the life-cycle damage of coal power plants, 4) (potentially if a relevant governmental policy is active) emission cost., i.e.,

$$\Phi = \sum_t \left[ 1^T y_t + 1^T y_t^\alpha + 1^T y_t^\beta + NC_t^T u_t + SU_t^T s_t + SD_t^T h_t + \Omega \left( Cost_\psi^T \psi + Cost_\xi^T \xi \right) \right]$$ where superscript $T$ represents a transpose.

V. 4. Proposed optimization problem

$$\min_{x_t^k} \sum_t \sum_k \left(1^T y_t^k + 1^T y_t^{k,\alpha} + 1^T y_t^{k,\beta} + NC_t^T u_t^k + SU_t^T s_t^k + SD_t^T h_t^k \right)$$

where $x_t^k = \left[\theta_t^k, g_t^k, g_t^{coalI,k}, g_t^{coalII,k}, u_t^k, u_t^{coalI,k}, u_t^{coalII,k}, s_t^k, h_t^k, \alpha_t^k, \beta_t^k, y_t^k, y_t^{k,\alpha}, y_t^{k,\beta}, v_t^k, \gamma_t^k, v_t^k, \delta_t^k \right]$

s.t.
1. $L_G \ g_t^k - B_{bus} \theta_t^k - d_t + LL\left(v_t^k - \gamma_t^k\right) = 0$
2. $\sum_t 1^T \left(L_G \ g_t^k\right) = \sum_t \left(1^T d_t\right)$
3. $g_t^{k,m} = g_t^{k,m,I} + g_t^{k,m,II} \quad m \in$ coal power plants
4. $-\text{line capacity} \leq B_{br} \theta_t^k \leq \text{line capacity}$
5. $-R \leq g_t^k - g_{t-1}^k \leq R$
6. $K_- u_t^k \leq g_t^k \leq K_+ u_t^k, \ K_-^{m,I} u_t^{k,m,I} \leq g_t^{k,m,I} \leq K_+^{m,I} u_t^{k,m,I}, \ K_-^{m,II} u_t^{k,m,II} \leq g_t^{k,m,II} \leq K_+^{m,II} u_t^{k,m,II}$
7. $g_{\max}^{m,I} u_t^{k,m,II} - g_t^{k,m,I} \leq 0 \quad m \in$ coal power plants
8. $u_t^k - u_{t-1}^k - s_t^k \leq 0, \quad u_{t-1}^k - u_t^k - h_t^k \leq 0$
9. $\sum_{j=t-\tau_s^i+1}^{t} s_j^{k,i} - u_t^{k,i} \leq 0, \quad \sum_{j=t-\tau_h^i+1}^{t} h_j^{k,i} \leq 1 - u_t^{k,i}$
10. $-\alpha_t^k + g_t^k - g_{t-1}^k \leq 0, \quad -\beta_t^k - g_t^k + g_{t-1}^k \leq 0$
11. $A_g g_t^k + B_g y_t^k \leq b_g, \ A_\alpha \alpha_t^k + B_\alpha y_t^{k,\alpha} \leq b_\alpha, \ A_\beta \beta_t^k + B_\beta y_t^{k,\beta} \leq b_\beta$
12. $\delta_{t+1,j} = \delta_{t,j} + \sum_k \text{Pr}_t^k \left[\varepsilon_{C\,j}\, \gamma_{t,j}^k - \left(\frac{1}{\varepsilon_{D\,j}}\right) v_{t,j}^k\right] \lambda_t \quad$ for $\forall\, t, k$
13. $v_{t,j}^k \lambda_t \leq \text{diag}\left(\varepsilon_{D\,j}\right) \delta_t^k \quad$ for $\forall\, t, j, k$
14. $\text{diag}\left(\varepsilon_{C\,j}\right) \gamma_{t,j}^k \lambda_t \leq \xi^k - \delta_t^i \quad$ for $\forall\, t, j, k$
15. $0 \leq v_{t,j}^k \leq \psi_j \quad$ for $\forall\, t, j, k$
16. $0 \leq \gamma_{t,j}^k \leq \psi_j \quad$ for $\forall\, t, j, k$
17. $0 \leq \delta_{t,j} \leq \xi_j \quad$ for $\forall\, t, j$
18. $\theta_t^k, g_t^k, \alpha_t^k, \beta_t^k, \gamma_{t,j}^k, v_{t,j}^k, \delta_t^k \geq 0$ and $u_t^k, u_t^{k,m,I}, u_t^{k,m,II}, s_t^k, h_t^k \in I$ in $\{0,1\}$

(10)

(10.1) is the real power balance equation; (10.2) is that the systemwide generation must be sufficient to serve all the loads over all the time periods; (10.3) represent that the power generation from a coal power plant is the sum of generation from Unit I and II; (10.4) is the flow limit; (10.5) is ramp up/down constraint; (10.6) is the operational constraints for each generator (max/min constraint) and each units; (10.7) links two independent generators that model a single coal power plant; (10.8) defines the start-up and the shut-down variables; (10.9) constrains minimum uptime and downtime; (10.10) defines the ramp up/down variables of coal power plants; (10.11) defines the feasible regions of the ramp up/down variables as shown in Fig. 6; (10.12) defines the charging state at the $(t+1)^{th}$ time slice in terms of multiple scenarios where $Pr^k$ stands for the probability that the scenario $k$ is realized; (10.13) and (10.14) are the limits of discharging and charging; (10.15) and (10.16) are the limits of the rates for charging and discharging; (10.17) represents the energy stored at a storage devices is at most its rated energy; and (10.18) identifies the variables where $I$ in $[0, I]$ is the integer between 0 and 1, i.e., either 0 or 1.

The optimization problems listed in (10) are an MILP that there are several commercial software packages available for solving. Because the newly added costs associated with impacts on environment and lifetime of the generator is monotonously increasing with ramp-up and ramp-down variables, the problems in (15) and (16) are convex. In this study, CPLEX is used for solving the problem.

## VI. SIMULATION RESULTS AND DISCUSSION

A modified IEEE-30 bus system shown in Figure 7 is the network for the simulation. Since the time frame of a unit commitment problem is a day, 24 hours periods are considered for the simulation. Generators 1, 2, and 3 are base units (Gens 1 and 2 are coal power plants while Gen 3 is a nuclear power plant), and Gen 4, 5, and 6 are peaking units, i.e., gas power plants. The generation capacities of all the generators are 60 MW except for the nuclear power plant Gen 3 (150 MW). The offers from the generators are summarized in Table I. Over 24-hour periods, the real power generation from Gen 3 barely changes due to its limited ramp rates. The ramp rates of the base units are [4 MW, 6 MW, 1 MW] and those of the peaking units are all 60 MW. Therefore, the peaking units can ramp up to their generation limits from cold start. In addition, the generation from the wind turbine located at Bus 12 is always fully dispatched, and the daily generation is peaked in the morning (during off-peak period of load) and set to zero in the rest of the day (blue lines for the wind resources from Figure 8). The EOL for the coal power plants (Gen 1 and 2) are the half of their generation limits, i.e., 30 MW. Even though the generation output from the wind farms follows a stochastic process, the output is considered as deterministic in this study because the decision to the UCP is made in a day ahead. However, the stochastic approach can be integrated by introducing multiple scenarios in terms of $k$ in (10). For the sake of simplicity, no storage devices are taken into consideration in these simulations.

It is possible to establish a piece-wise-linear cost curve with $\alpha_t$ and $\beta_t$ in a similar way that MATPOWER considers the block-type offer curve [19]. In this study, we assume a simple linear curve for the additional costs. the proposed carbon tax is in a wide range; \$10/tCO$_2$e according to the report from the World Bank [20], but a study with €30/tCO$_2$ (about \$33/tCO$_2$) may not be sufficient to reduce the emission [21]. In addition to the carbon tax, the cost associated with the internal damage due to the deep cycling is assumed in the range of $1 - 10$ times greater than that of carbon tax. In the simulation, for comparison, various costs are taken into consideration associated with $\alpha_t$ and $\beta_t$ are considered; 1) zeros, [0; 0], 2) low, [15; 8], 3) high, [150; 80], and 4) very high [450; 240] \$/MW. The scenario with zero cost for $\alpha_t$ and $\beta_t$ is like the current electricity markets in the US where the unit commitment decision is left to the generator owners. The low and the high cost scenarios are the power industries in the consideration of the carbon tax and the internal damage. The very high cost scenario is highly conservative way to minimize the deep cycling. To examine the impact of variable energy resources, we performed the same simulation with and without the wind turbine. While both the load and the wind output change over time, the change of wind is much more drastic than that of the loads which forces the coal power plants to cycle.

Figure 9 illustrates the simulation results in terms of the additional costs associated with $\alpha_t$ and $\beta_t$ without the integration of wind turbines. Over 24 hour period, the load changes and the generation dispatch and unit commitment changes to meet the varying loads. From the top to the bottom row, the marginal cost with $\alpha_t$ and $\beta_t$ increases as was described. The left column presents the generation dispatches and unit commitments, and the right column shows the detailed dispatch from Unit I and II for the coal power plants (Gen 1 and 2). In the current US electricity markets, the additional costs are not taken into consideration as mentioned above. The first row indicates the simulation results of the current US market. The dispatches from all the power plants except Gen 3 and 6 changes to meet the fluctuating loads. In the simulation, the offers from Gen 6 are too expensive to get dispatched. While the offers from Gen 3 is least expensive, the ramp rate limits for the generator are highly constraining, which makes the dispatch from the generator almost fixed during the simulation. At the peaking times, both coal and gas power plants follow the temporal changes in the load. The dispatches from coal power plants at $t = 0$ is low, and the ramp rates for the coal power plants are not large enough to catch the load changes. Therefore, to meet the load before the peaking time, the gas power plants are necessary. However, after the peaking time, the coal power plants are enough to meet the loads, which is why the gas power plants are turned off. As the addition costs increase (from the top to the bottom rows), the temporal variations of coal power plants below the EOL become decreased. The change in the dispatch pattern of the coal power plants does not affect the dispatch of other generators (Gen 3, 4, 5, and 6) significantly because the temporal change of the load is smooth in time such that the minor change in the dispatch is sufficient to compensate the change. In other words, the integration of the additional cost does not change the operation of the electricity markets significantly in terms of dispatch. This may shed light on why the current market structure with limited integration of variable energy resources does not seek for a new tool to accommodate the addition cost related to $α_t$ and $β_t$.

As the stochastic energy resources are integrated into the electric power industry, the temporal changes can be significant. Figure 8 shows the wind resource at Bus 12, which was not utilized in the simulation of which the result is shown in Figure 9. As shown in Figure 8, the wind energy is small in comparison to total loads, but the temporal changes of the wind resources are much grater than those of total loads. Figure 10 illustrates the results of simulation with the wind turbines. Note that the only difference in the simulations in Figure 9 and 10 is whether to utilize the wind energy. At the top row, the results with no additional costs with $α_t$ and $β_t$ are presented, which is similar to those without the wind turbine (See the top row in Figure 9). As the marginal costs associated with $α_t$ and $β_t$ become increased, the dispatches from the Unit I's of the coal power plants tend fixed. As a result, the integration of the additional cost prevents the deep cycling of coal power plants (the changes in dispatch below EOL). Different from the results shown in Figure 9, the changes in dispatches from Unit I's of the coal power plants are very large so that the other generators must adjust the dispatch as well. It is interesting to note that the gas power plants with the highest operation cost (Gen 6, black dotted line in Figure 10) are optimally dispatched when the wind turbine must be dispatched, and the additional cost associated with $α_t$ and $β_t$ are very high.

## VII. CONCLUSIONS

The $CO_2$ emission from coal power plants is analyzed in terms of ramping (the temporal change in the generation output) as well as the generation output itself. From the real-world example, the dynamic emission is evident below a certain level (deep cycling). This study suggests that the damage by deep cycling be introduced in the operation cost for economic studies and the dynamic emission be included to address the subject of emission and the internal damage of the coal power plants. A unit commitment problem is formulated to assess the impact of government policy to reduce the greenhouse gas emission as well as the reduced lifetime of the coal power plants. The simulation results indicate that carbon tax effectively tailors the generation patterns of polluting generators to reduce the greenhouse gas emission and the intertemporal ramping of coal power plants.

## ACKNOWLEDGEMENT

The author appreciates *Xcel Energy* for sharing the greenhouse gas emission data.

Table I. The offers from the generators; $q$ is in MW and $p$ is in $/MWh

|       | Gen 1 | | Gen 2 | | Gen 3 | | Gen 4 | | Gen 5 | | Gen 6 | |
|-------|----|----|----|----|-----|-----|----|-----|----|-----|----|-----|
| Block | $q$ | $p$ | $q$ | $p$ | $q$ | $P$ | $q$ | $p$ | $q$ | $p$ | $q$ | $p$ |
| 1     | 25 | 7  | 25 | 5  | 25  | 1   | 25 | 50  | 25 | 80  | 25 | 120 |
| 2     | 20 | 10 | 20 | 12 | 20  | 4   | 20 | 100 | 20 | 300 | 20 | 750 |
| 3     | 15 | 15 | 15 | 20 | 105 | 4.7 | 15 | 150 | 15 | 500 | 15 | 999 |

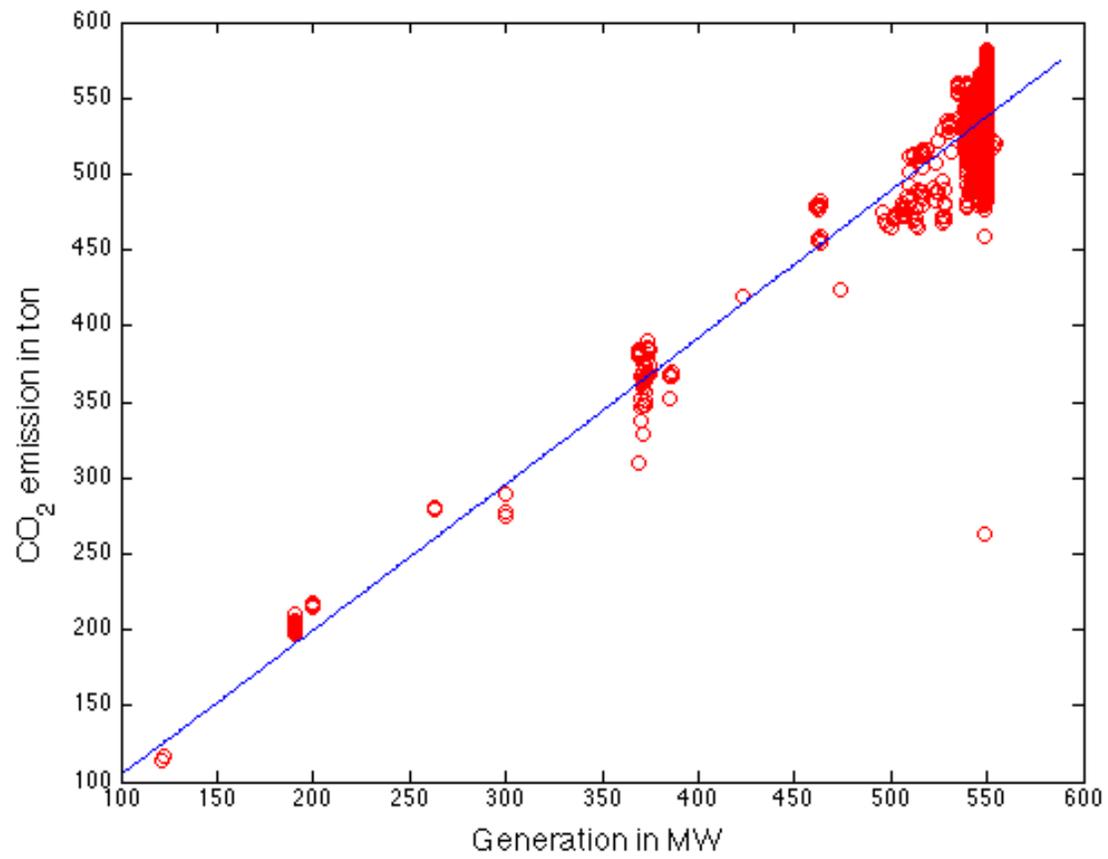

Figure 1. Fitting results of measured static $CO_2$ emission as a function of generation output to (3).

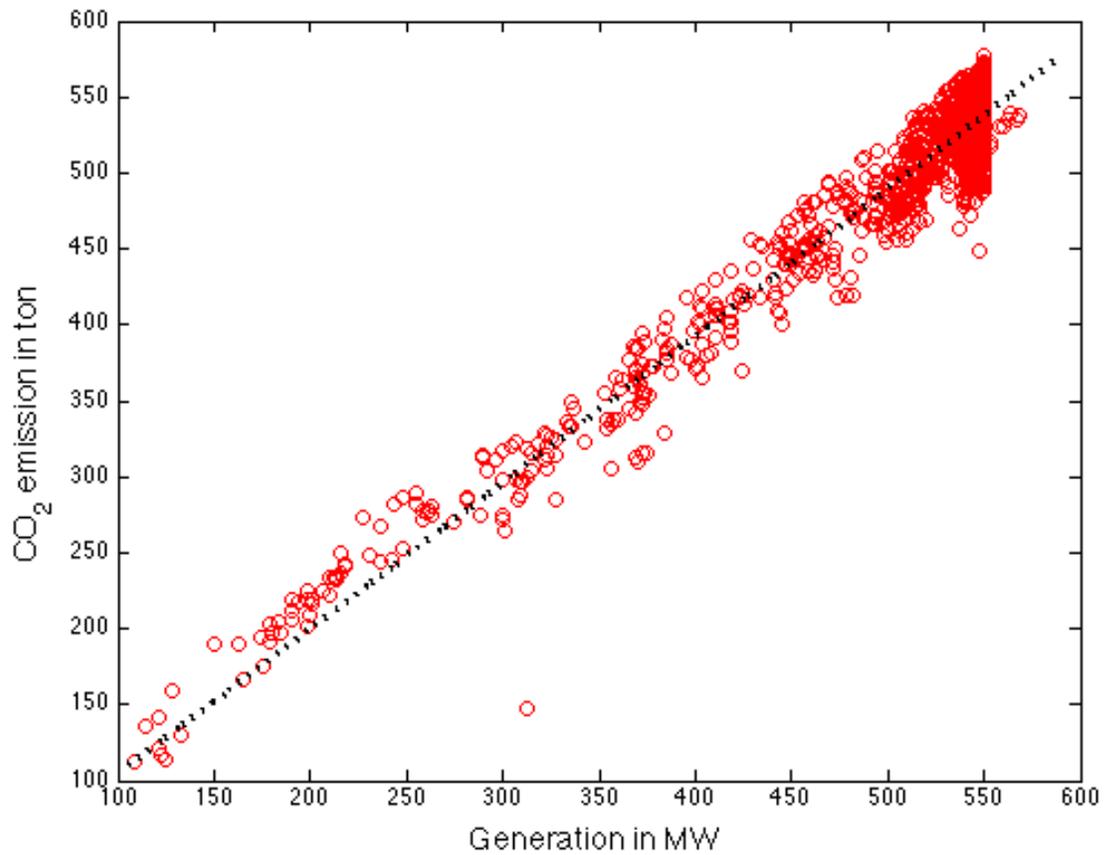

Figure 2. Hourly $CO_2$ emission data of a coal power plant that undergoes the change in its output. The dotted line indicates $CO_2$ emission if no dynamic effect is present. Above the half of the maximum generation, the data are evenly spread around the dotted line. On the other hand, below 300MW (about half of the maximum generation), the actual $CO_2$ emission data stay above the dotted line indicating that the dynamic effect is evident in the range.

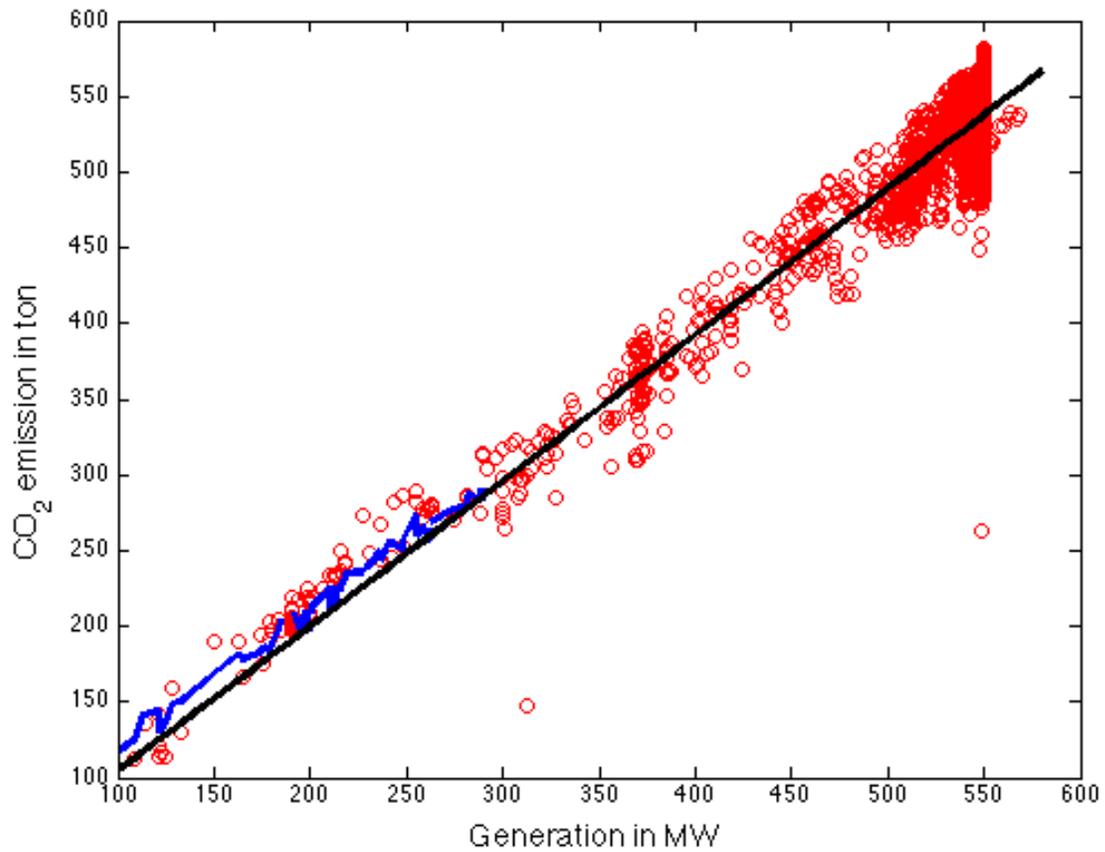

Figure 3. Fitting results of $CO_2$ emission to (6). The dotted line is an estimated emission based on the linear model by ignoring the dynamic effect, and the blue line indicates the line representing (6) with $\tau = 0.34 \pm 0.27$, $b = 6.12 \pm 2.55$, and $N_2 = 0.20 \pm 0.10$.

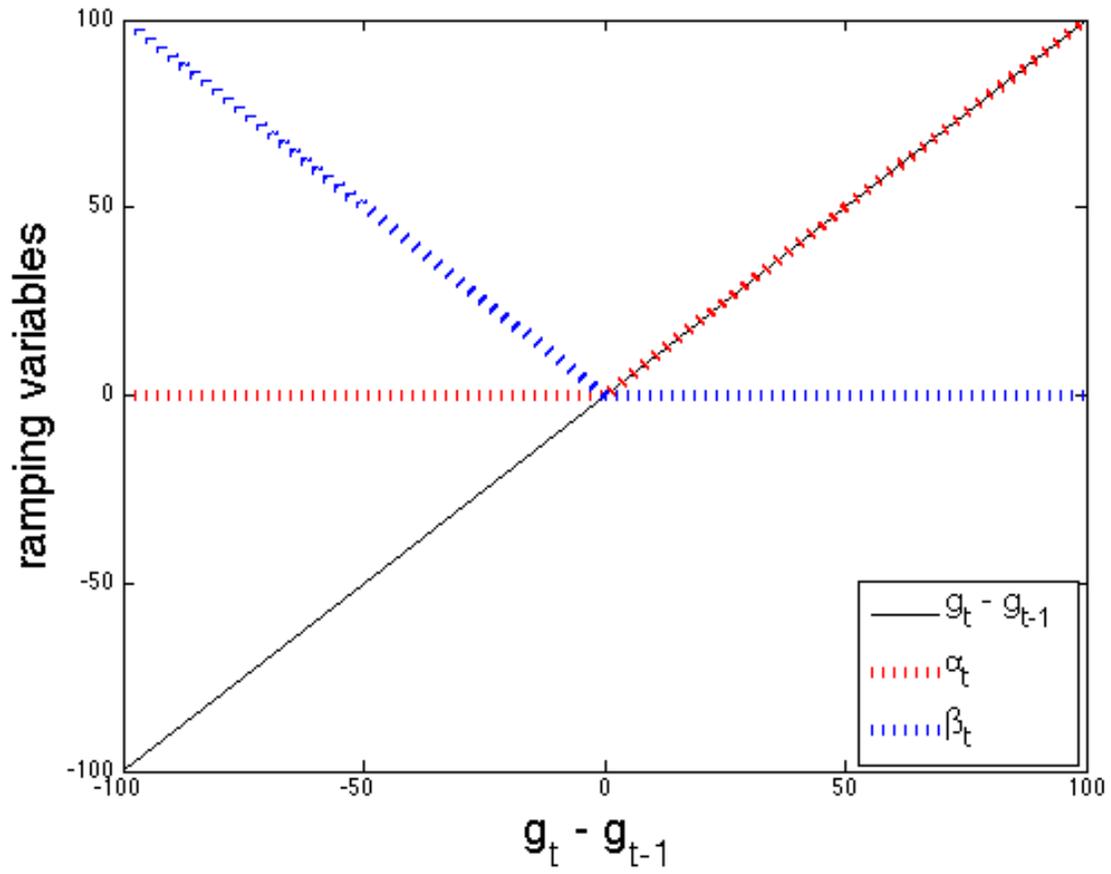

Figure 4. The definition of ramping variables $α_t$ and $β_t$ in terms of the change in generation output.

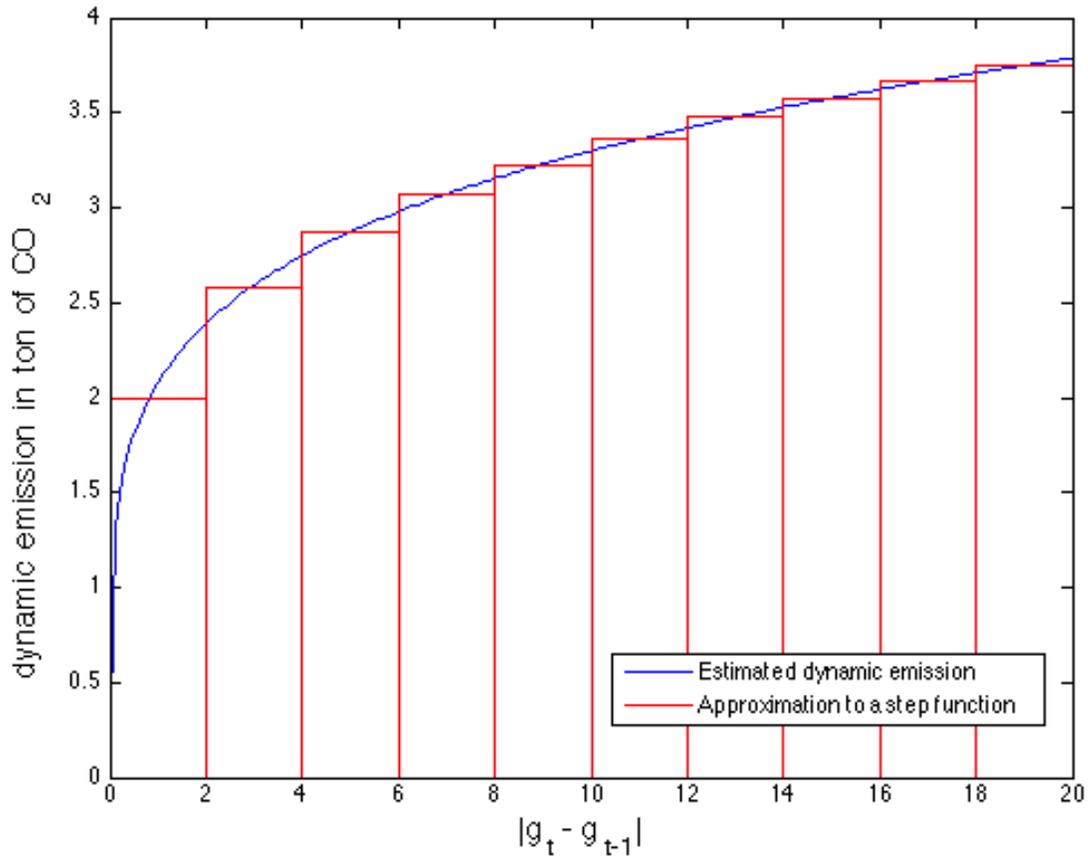

Figure 5. The estimated dynamic emission of $CO_2$ with respect to the ramp rate (blue line). The red line represents the approximation of the dynamic emission with a step function so that the areas under blue and red lines are maintained same. Note that the horizontal axis refers either $\alpha$ (ramp up; $\alpha > 0$, $\beta = 0$), or $\beta$ (ramp down; $\alpha = 0$, $\beta > 0$).

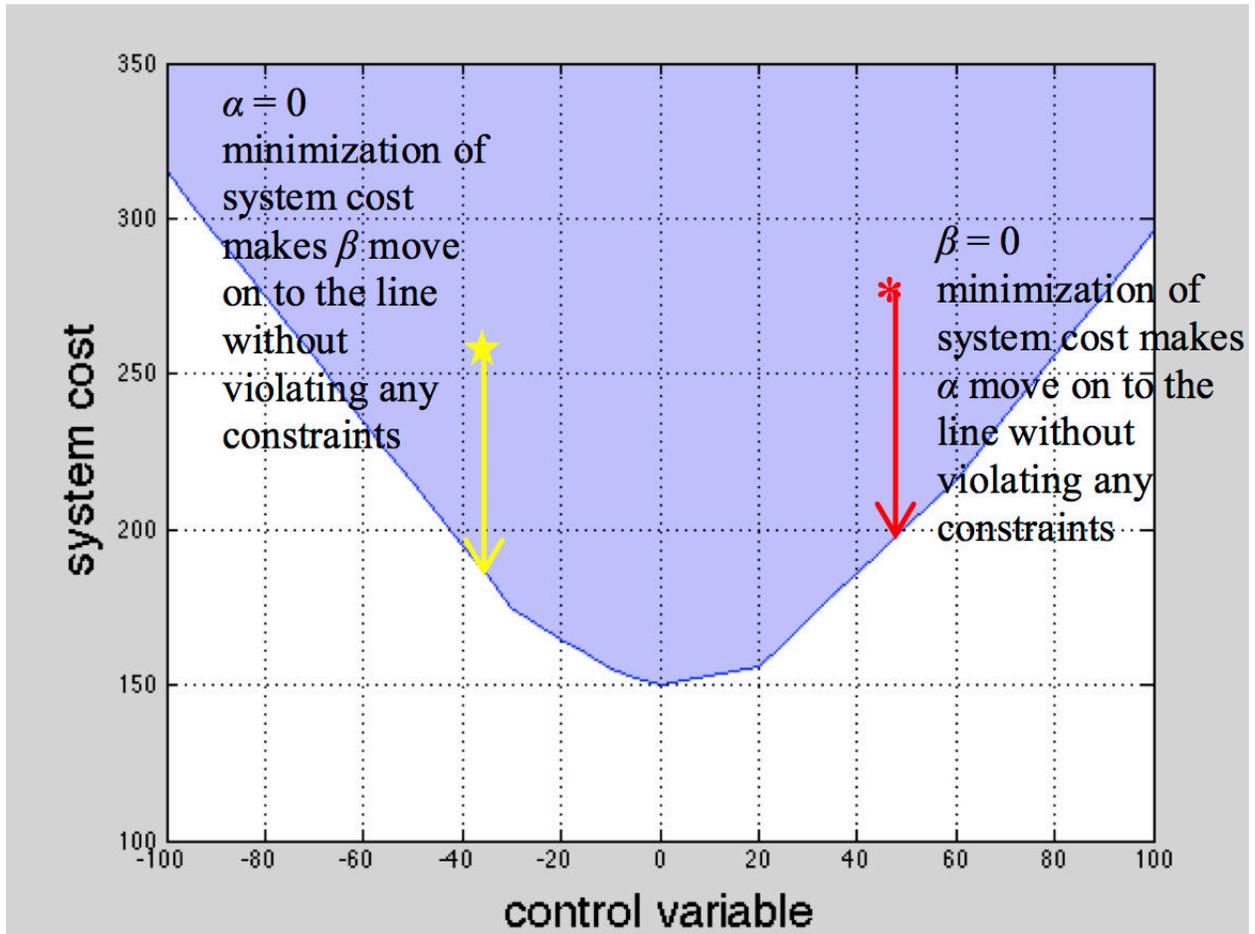

Figure 6. Graphical illustration how cost minimization process efficiently makes inequality constraints (10.10) identical to equality constraints (8).

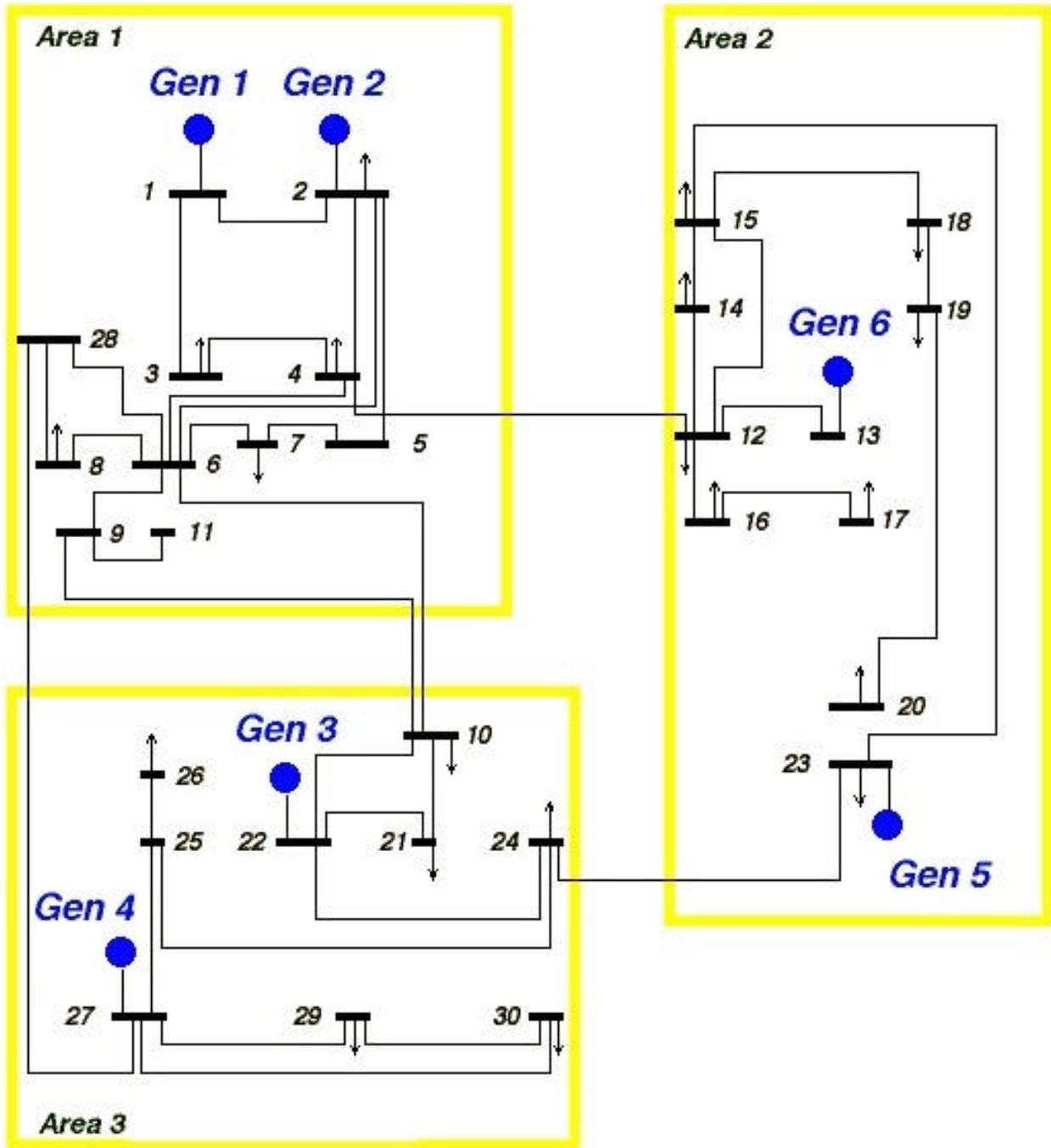

Figure 7. Modified IEEE 30 bus system; Generator 1 at Bus 1 and Generator 2 at Bus 2 are coal power plants, and a wind generator is located at Bus 12.

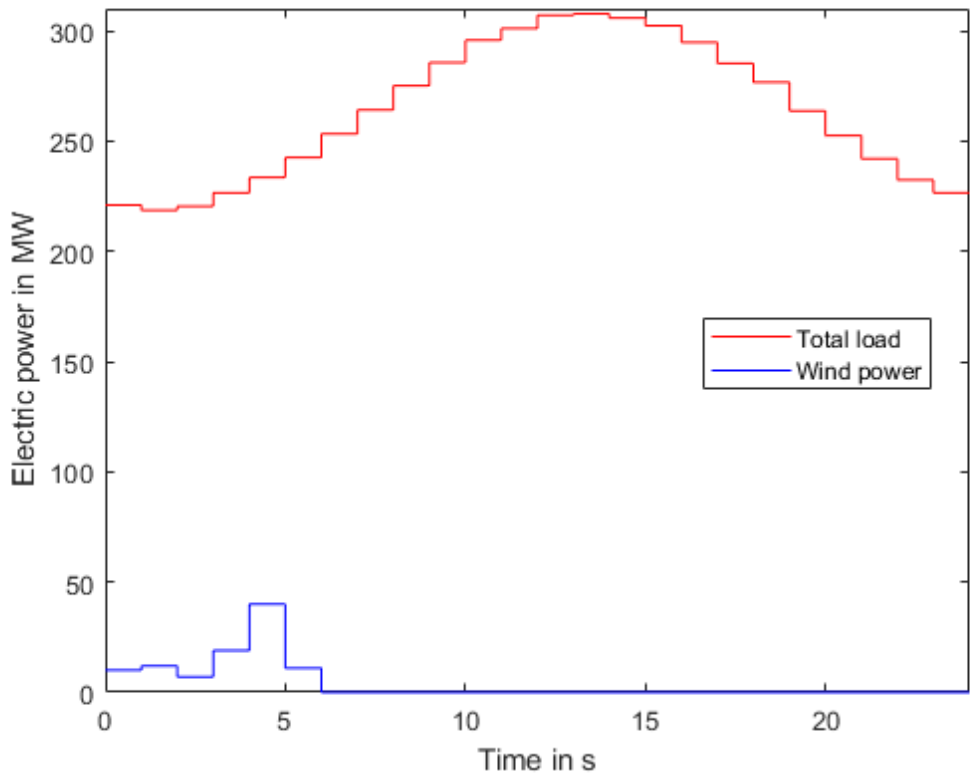

Figure 8. The variation of total loads and of the wind power resources over a day (24 hrs).

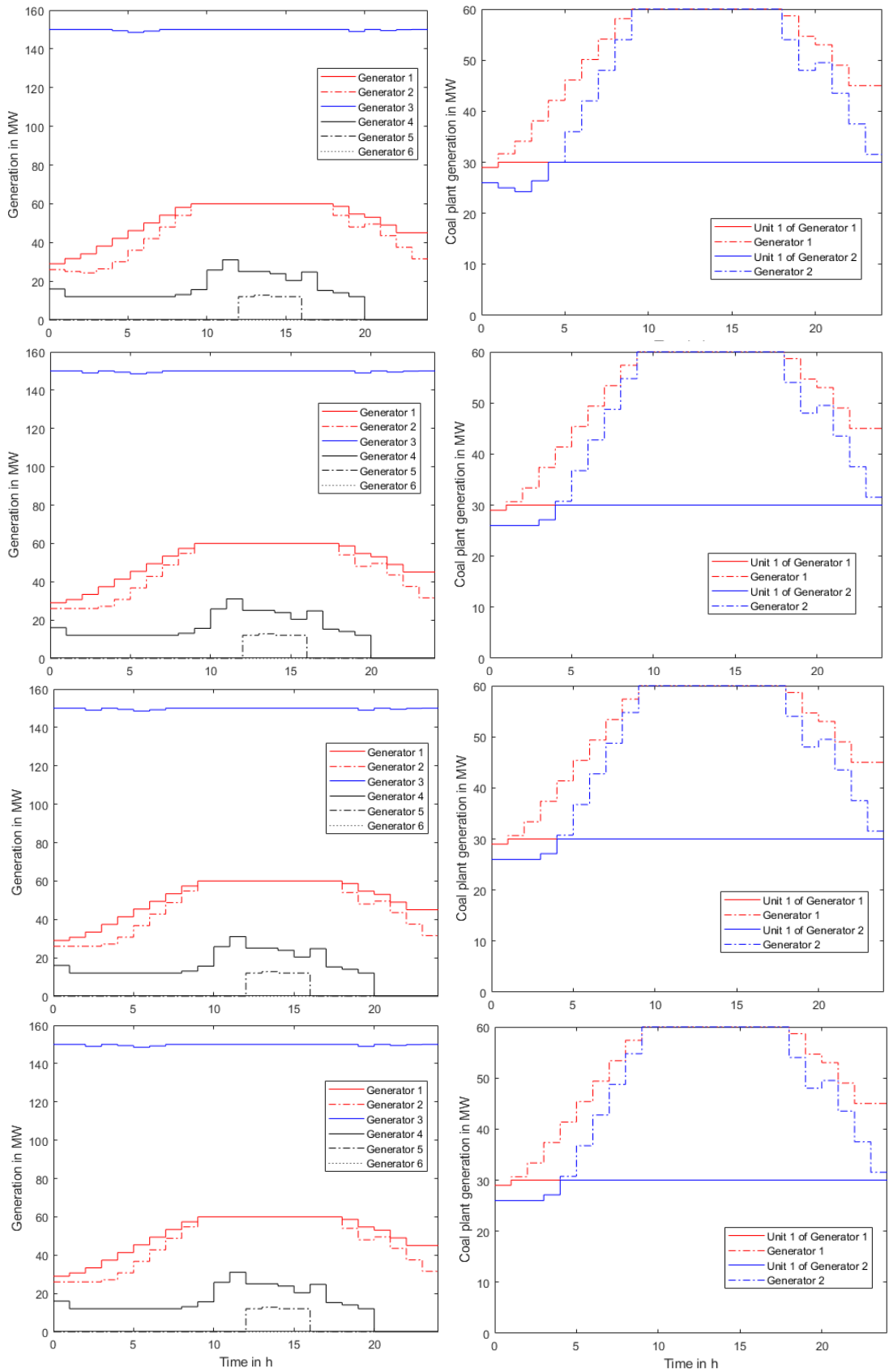

Figure 9. Generation outputs (left column) and generation from coal power plants (right column) with no wind turbine. Costs associated with $α_t$ and $β_t$ are zeros (top), low (2$^{nd}$ row), high (3$^{rd}$ row) and very high (bottom row).

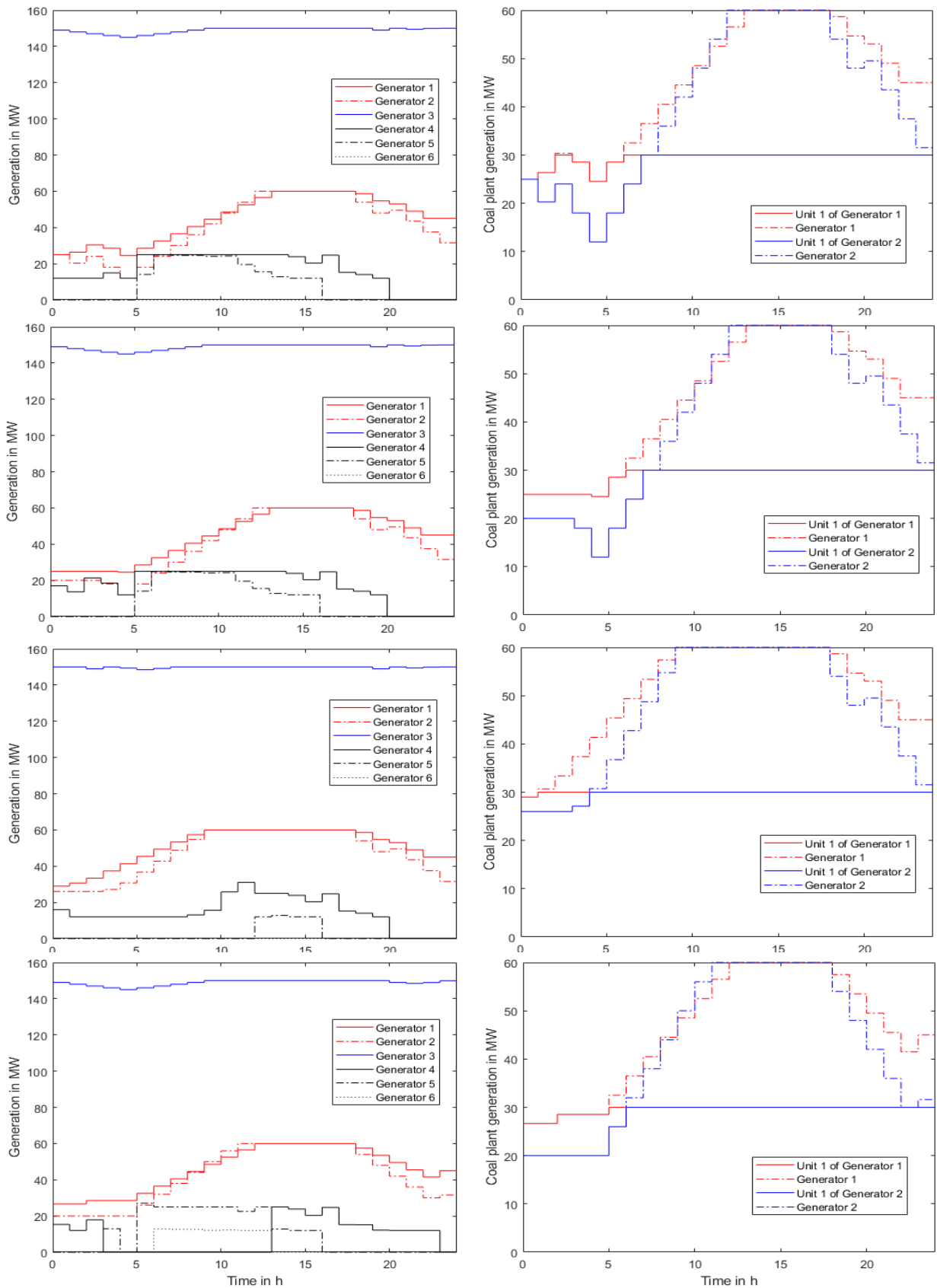

Figure 10. Generation outputs (left column) and generation from coal power plants (right column) with wind turbine at Bus 12. Costs associated with $\alpha_t$ and $\beta_t$ are zeros (top), low (2$^{nd}$ row), high (3$^{rd}$ row) and very high (bottom row).